\documentclass[useAMS,usenatbib]{mn2e}

\usepackage{graphicx}
\usepackage{epstopdf}
\usepackage{widetext}
\usepackage[bottom,flushmargin,hang]{footmisc}
\usepackage[fleqn]{amsmath}
\usepackage{amssymb}
\usepackage{subfigure}
\usepackage{url}
\usepackage{caption}
 
\newcommand\beq{\begin{equation}}
\newcommand\eeq{\end{equation}}
\newcommand\beqar{\begin{eqnarray}}
\newcommand\eeqar{\end{eqnarray}}

\newcommand{\mycite}[2]{\citeauthor{#1}~(\citeyear{#1};~\citealp[see also][]{#2})}

\title[Probing IGMF with EGB Anisotropy]{Probing the Intergalactic Magnetic Field with the Anisotropy of the Extragalactic Gamma-ray Background}
\author[T. M. Venters and V. Pavlidou]{T. M. Venters$^{1}$\thanks{E-mail:
tonia.m.venters@nasa.gov; pavlidou@physics.uoc.gr} and V.
Pavlidou$^{2,3}$\\
$^{1}$Astrophysics Science Division, NASA Goddard Space Flight Center, Greenbelt, MD, 20771, USA\\
$^{2}$Max-Planck-Institut-f\"{u}r-Radioastronomie, Auf dem H\"{u}gel 69, DE-53121, Bonn, Germany\\
$^{3}$Department of Physics, University of Crete, 71003 Heraklion, Greece}
\begin{document}


\pagerange{\pageref{firstpage}--\pageref{lastpage}} \pubyear{2013}

\maketitle

\label{firstpage}

\begin{abstract}
The intergalactic magnetic field (IGMF) may leave an imprint on the angular anisotropy of the extragalactic gamma-ray background through its effect on electromagnetic cascades triggered by interactions between very high energy photons and the extragalactic background light. A strong IGMF will deflect secondary particles produced in these cascades and will thus tend to isotropize lower energy cascade photons, thereby inducing a modulation in the anisotropy energy spectrum of the gamma-ray background. Here we present a simple, proof-of-concept calculation of the magnitude of this effect and demonstrate that current {\it Fermi} data already seem to prefer non-negligible IGMF values.
The anisotropy energy spectrum of the {\it Fermi} gamma-ray background could thus be used as a probe of the IGMF strength.
\end{abstract}


\section{Introduction}\label{sec:intro}

Cosmic magnetic fields are expected to play a fundamental role in the physics of a large variety of astrophysical systems. The fate of high energy particles, such as ultra-high energy cosmic rays (UHECRs), as they propagate through the Universe and the acceleration of charged particles in astrophysical objects hinge, in large part, on magnetic fields within their sources and those permeating the Universe. Large-scale magnetic fields such as those found in galaxies and clusters of galaxies have been a mystery for the past several decades. While they are generally thought to be the result of the amplification of weak seed fields, the origins of these seed fields, whether they be cosmological or astrophysical in nature, are largely unknown. A definitive measurement of the intergalactic magnetic field (IGMF) could provide a fundamental step in resolving the questions of the origins of cosmic magnetic fields and their impact on the evolution of the systems in which they reside, but sufficiently constraining observations have thus far remained elusive. Previously, upper bounds of $\sim 10^{-9}$ G have been found through Faraday rotation limits of polarized radio emission from distant quasars and the study of the effect of magnetic fields on the anisotropy of the cosmic microwave background \citep[for a brief review of observational and theoretical techniques to constrain the IGMF, see e.g.,][and references therein]{ner09}. The recent availability of data from the {\it Fermi} Large Area Telescope (LAT) has renewed interest in the IGMF as observations of the gamma-ray sky and its participating source populations provide a unique perspective on the strength and structure of the IGMF.

The gamma-ray sky consists of resolved point sources (such as normal and active galaxies, pulsars, etc.), transient gamma-ray sources (e.g., gamma-ray bursts), and the diffuse gamma-ray radiation comprised of emission from the Galaxy and the isotropic (presumably, extragalactic) gamma-ray background (EGB). The EGB is a window into the high-energy processes in the Universe, and its origins have been the subject of much debate. It is expected that emission arising from unresolved, extragalactic point sources, such as blazars and star-forming galaxies (SF), comprises a sizable contribution to the EGB \citep[see e.g., ][ and references therein]{ss96a,fie10,sclat10,sv11}. Additionally, many of these extragalactic point sources are also sources of very high energy (VHE) gamma rays,\footnote{In this paper, we take VHE to be $\sim$ TeV.} which interact with the soft photons of the extragalactic background light (EBL), consisting of the cosmic microwave background (CMB) and the background of infrared, optical, and ultraviolet radiation from direct and dust-reprocessed starlight. The interactions of VHE photons with the EBL initiate electromagnetic (EM) cascades, giving rise to another contribution to the EGB \citep[see e.g.,][]{cop97,kne08,ino09,ven10}. However, the degree to which astrophysical sources contribute to the EGB depends on their distribution with respect to luminosity and redshift (the gamma-ray luminosity function, GLF), their intrinsic spectra at gamma-ray energies, and the nature of the EBL, all of which remain the subjects of intense debate \citep[see e.g.,][]{ino09,ven10,fie10,ebllat10,sclat10,mak11,ino11,sv11,aba11}.  

A promising complementary technique for investigating the contributions of astrophysical sources to the EGB involves studying the angular fluctuations in intensity of the EGB as a function of energy, the anisotropy energy spectrum. As pointed out in \mycite{sp09}{hen10,sie11}, the contribution of a given class of sources to the overall fluctuation angular power at a given angular scale of the EGB at a given energy is weighted by the square of its fractional contribution to the intensity of the EGB at that energy. As such, if the relative contributions of the various EGB contributors change with energy, so too will the fluctuation angular power as a function of energy. Expectations for the anisotropies of the commonly invoked astrophysical contributors to the EGB are that SF galaxies contribute very little anisotropy owing to the fact that they are so numerous and individually faint, while blazars, being far more rare than SF galaxies but much brighter, could have a large contribution to the EGB anisotropy \citep{and07,ap09,sp09,hen10,anisolat12}. It has also been suggested that annihilation in dark matter substructures in the Galactic halo could also have a large contribution to the EGB anisotropy, even if the contribution to the EGB intensity from dark matter annihilation is subdominant with respect to other contributions \citep[see e.g.,][]{and07,sie08,tao09,and09,sp09,hen10}. Early anisotropy results from the {\it Fermi}-LAT Collaboration \citep{anisolat12} suggest that the energy dependence of the fluctuation angular power of the EGB may be consistent with arising from a source population with spectra similar to that of blazars ($\alpha \sim 2.4$), even though blazars may comprise $\lesssim 30\%$ of the EGB intensity \citep{sclat10,cuo12}. Thus, even contributions to the EGB that are subdominant with respect to other contributions can have a profound impact on the anisotropy of the EGB, especially if the relative prominence of its components changes with energy. As such, the anisotropy energy spectrum could prove a powerful tool in detecting changes in the EGB anisotropy resulting from the magnetic deflection of EM cascades from blazars.

In interacting with EBL photons, VHE gamma rays initiate EM cascades of photons, electrons, and positrons. The interaction of a VHE gamma ray with an EBL photon produces a pair of electrons and positrons, which will, in turn, Inverse Compton (IC) scatter EBL photons to high energies. These up-scattered photons will, in turn, pair produce, and the process continues until the energies of the resulting photons are low enough that pair production is no longer efficient. In the presence of an IGMF, the charged particles in the cascade are deflected away from the direction of propagation of the primary photon. As these charged particles propagate through the EBL, they can up-scatter soft photons towards the observer, though they will appear to be coming from a different direction from the GeV gamma rays. Observationally, this process results in a ``halo'' of gamma rays around the source. The detection of a halo around a gamma-ray point source would not only provide strong evidence for an IGMF, but would also provide an indication of the strength of the IGMF. Several groups have already conducted searches for extended emission around gamma-ray sources \citep[see e.g.,][]{aha01,ner10,and10,ale10,ner11,arl12}, though as yet, no significant indication of gamma-ray halos distinguishable from instrumental effects have been reported. However, the absence of significant halo emission could imply a strong IGMF that would extend the EM cascade halo such that the surface brightness of the extended source falls below the detection threshold \citep{ner10,and10}. In addition to possibly creating a halo, the deflection of EM cascades by the IGMF could result in a reduction in observable cascade radiation. Analyses employing simultaneous observations of blazars by atmospheric Cherenkov telescopes and {\it Fermi} make use of this fact to derive limits on the IGMF \citep{der11,ess11,tay11,kac11}. These techniques are quite promising and, with more data from {\it Fermi}, will continue to shed light on the IGMF. In this paper, we consider a different promising technique that makes use of the anisotropy energy spectrum of the EGB to constrain the IGMF. \\
\indent For any cosmological population, such as blazars, that emits gamma rays at very high energies, the effect of EM cascading results in a flux suppression at the highest energies and enhancement at lower energies as the cascades redistribute radiation from the higher energies to the lower energies \citep[see e.g.,][]{str73,str74,cop97,kne08,ino09,ven10}. In considering the spectral subpopulations of blazars, flat-spectrum radio quasars (FSRQs) and BL Lacertae-type objects (BL Lacs), the intrinsic spectra of FSRQs are much softer than that of both BL Lacs and EM cascades, and their relative prominence with respect to the EGB can change with energy \citep{ven10,vp11}. As such, the fluctuation angular power in the EGB should change with energy \emph{even if FSRQs and BL Lacs have similar anisotropy properties.} Furthermore, as noted earlier, the deflection of EM cascades by the IGMF can spatially extend the emission around gamma-ray sources and reduce the observable cascade radiation. Thus, the impact of a strong IGMF is to \emph{reduce} the blazar anisotropy both because of the increased angular scale of the sources and because of the reduction in the overall blazar contribution to the EGB (assuming that blazars are the most anisotropic EGB component). On the other hand, in the absence of a strong IGMF, the EM cascades are highly collimated in the direction of the primary photon (angular spread $\sim m_e/E_e$), and their impact on the blazar anisotropy is to \emph{reinforce} it due to the enhancement in the overall blazar contribution to the EGB. Hence, in the two extreme cases for the IGMF, EM cascades could have a profound and \emph{opposite} impact on the blazar anisotropy and, by extension, the EGB anisotropy. Therefore, studies of the EGB anisotropy could prove a sensitive probe into the nature of the IGMF.

In this paper, we explore the impact of EM cascade radiation from blazars on the EGB anisotropy in the extreme cases of a negligible IGMF and an IGMF that is sufficiently strong to essentially isotropize the cascade emission (i.e., from the \emph{observationally} practical standpoint). This study is meant to serve as a proof of concept for the feasibility of using EGB anisotropies as an IGMF probe. In Section \ref{sec:method}, we present the methodology and parameters employed to determine the anisotropy energy spectrum and the EM cascades. In Section \ref{sec:results}, we present the results of the calculation. 

\section{Methodology}\label{sec:method}

\subsection{Collective Intensity of Unresolved Astrophysical Sources}\label{subsec:blazars}

The contribution to the EGB due to unresolved blazars can be viewed as the superposition of the collective intensity of \emph{intrinsic} blazar spectra and the contribution from the cascade radiation from the interactions of VHE photons from blazars with the EBL:
\beq
I^{\rm bl}_E(E_0) = I^i_E(E_0) + I^c_E(E_0)\,,
\eeq
where the intensity, $I_E(E_0)$, is given in units of photons per unit area per unit time per unit solid angle per unit energy emitted at observer frame energy $E_0$. In calculating the collective spectrum of unresolved blazars, we follow the formalism outlined in \citet{vp11}. Blazar spectra are taken to be smoothly broken power laws:
\beq
F_E(E_0) = F_0\left[\left(\frac{E_0}{E_b}\right)^{\alpha_1n}+\left(\frac{E_0}{E_b}\right)^{\alpha_2n}\right]^{-1/n},
\eeq
where $F_E(E_0)$ is the differential photon flux in units of photons per unit area per unit energy per unit time, $E_b$ is the break energy, $\alpha_1 = \alpha - \Delta\alpha_1$ is the low-energy slope, $\alpha_2 = \alpha + \Delta\alpha_2$ is the high-energy slope, $\alpha$ is the gamma-ray photon spectral index of the blazar assuming an unbroken power-law spectrum,\footnotemark \ and $n$ quantifies the sharpness of the transition from the low-energy power law to the high-energy power law (taken to be $1$). 

\footnotetext[2]{For the sake of clarity, we use $\alpha$ for the photon index rather than $\Gamma$, which is commonly used in the literature.}

The total flux, $F$, of photons with energies greater than some fiducial energy,  $E_f$ (taken to be $100$ MeV), is found by integrating $F_E(E_0)$ over energy,
\beq
F = F_0\int_{E_f}^{\infty}\left[\left(\frac{E_0}{E_b}\right)^{\alpha_1}+\left(\frac{E_0}{E_b}\right)^{\alpha_2}\right]^{-1}dE_0\,.
\eeq
Then, the contribution of a single unresolved blazar to the EGB is
\beq\label{eqn:contoneblazar}
I_1 = \frac{F\left[(E_0/E_b)^{\alpha_1}+(E_0/E_b)^{\alpha_2}\right]^{-1}}{4\pi\int_{E_f}^{\infty}\left[(E_0/E_b)^{\alpha_1}+(E_0/E_b)^{\alpha_2}\right]^{-1}dE_0} 
\eeq
where the flux of one source is uniformly distributed over the entire sky in anticipation of an isotropically distributed cosmological population. Including absorption due to the EBL, the contribution becomes
\beq\label{eqn:contoneblazarebl}
I_1 = \frac{F\left[(E_0/E_b)^{\alpha_1}+(E_0/E_b)^{\alpha_2}\right]^{-1}}{4\pi\int_{E_f}^{\infty}\left[(E_0/E_b)^{\alpha_1}+(E_0/E_b)^{\alpha_2}\right]^{-1}dE_0} e^{-\tau(E_0,z)} \,
\eeq
where $\tau(E_0,z)$ is the optical depth for a photon with observer frame energy $E_0$, emitted at redshift $z$. In determining the total unresolved blazar contribution including EBL absorption, one would generally rewrite Equation \ref{eqn:contoneblazarebl} in terms of the blazar gamma-ray luminosity and redshift, multiply by the blazar GLF and \emph{intrinsic} spectral index distribution (corrected for the spectral bias endemic to flux-limited surveys\footnotemark\ as per \citet{vpr09}), and finally integrate over gamma-ray luminosity, redshift, and spectral index, thereby accounting for evolution in both the sources and the EBL (through the redshift-dependent optical depth; see e.g., \citealt{vpr09}). However, in performing a calculation that includes cascades, the population of blazars must be divided into its subpopulations of FSRQs and BL Lacs since these subpopulations have distinct spectral properties. In order to perform a full calculation that accounts for the evolution of sources (as in \citet{ven10}), we would require reliable GLFs for each subpopulation. In the case of FSRQs, a reliable GLF does exist~\citep{aje12}, but one has yet to be determined for BL Lacs since it is difficult to obtain reliable redshifts for them. While BL Lac GLFs based on EGRET and early {\it Fermi} data do exist \citep[see e.g.,][]{ino09}, a BL Lac GLF based on the current and more sensitive {\it Fermi} dataset remains in the works~\citep{aje12aas}, and it is as yet unclear how the possible modification of BL Lac spectra by cascades could impact the interpretation of a LAT-measured BL Lac GLF (though, work on this topic is currently being performed by the authors). Given that the purpose of this paper is to serve as a proof-of-concept for using relative changes in the EGB anisotropy to constrain the IGMF, for the sake of simplicity, we do not incorporate full blazar GLFs in our calculations. Rather, we assume typical luminosities and redshifts for all blazars, deferring a more detailed calculation to the future, at which time reliable GLFs for both FSRQs and BL Lacs will have been measured.
%
As such, the only integral that must be performed is that over the spectral index:
\footnotetext[3]{We note that the {\it Fermi} survey is not exactly a flux-limited survey due to the non-uniformity of the total diffuse background throughout the sky. However, for the purposes of this paper, we neglect this effect.}
\begin{multline}\label{eqn:EGB-decoupled}
I^{\rm bl}_E(E_0) = \\ 
I_0\!\! \int_{-\infty}^{\infty} \!\!\!\!\!\!\! d\alpha \, p(\alpha)\frac{\left[(E_0/E_b)^{\alpha_1}+(E_0/E_b)^{\alpha_2}\right]^{-1}}{\mathcal{S}(E_f,\alpha)}e^{-\tau(E_0,z)}\,,
\end{multline}
where $I_0$ is a normalization constant, $p(\alpha)$ is the \emph{measured} spectral index distribution (SID), and
\beq
\mathcal{S}(E_f,\alpha) = 
\int_{E_f}^{\infty} \!\!\! dE_0 \left[\left(\frac{E_0}{E_b}\right)^{\alpha_1}+\left(\frac{E_0}{E_b}\right)^{\alpha_2}\right]^{-1}.
\eeq
We fix the normalization constant so that the total astrophysical (blazars, SF galaxies, and EM cascades) contribution to the EGB fits the spectrum of the EGB as measured by the {\it Fermi}-LAT \citep{lat10a}. 

In determining the measured SID, we make use of the results presented in \citet{vp11} from a likelihood analysis fitting {\it Fermi}-LAT First Catalog \citep{agnlat10} FSRQs and BL Lacs to Gaussian SIDs accounting for errors in measurement of individual blazar spectral indices. For these blazar subpopulations, the maximum-likelihood Gaussian SIDs can be characterized by means ($\alpha_0$) and spreads ($\sigma_0$) with parameters determined to be $\alpha_0 = 2.45$ and $\sigma_0 = 0.16$ for FSRQs and $\alpha_0 = 2.17$ and $\sigma_0 = 0.23$ for BL Lacs. Following \citet{vp11}, we model the spectral breaks by taking $\Delta\alpha_1 = 0.1$, $\Delta\alpha_2 = 0.9$, and $E_{b,0} = 4$ GeV for FSRQs, though we take $E_{b,0} = 10$ TeV for BL Lacs \citep[treating them as High-frequency--peaked BL Lacs, which do not exhibit breaks in the LAT energy range; see e.g., ][]{spclat10}.

In principle, one would prefer to perform the above procedure on subsamples of local universe blazars to ensure that only \textit{intrinsic} spectra are used (i.e., that the spectra have not already been modified by EBL absorption and EM cascades). However, the population of local universe sources is sparse, and the results of a likelihood analysis performed on such a population would be subject to a considerable degree of uncertainty. Even so, in the case of FSRQs, the spectra are too soft to produce an appreciable amount cascade radiation \citep{ven10}. For this reason and because the photon index of the spectrum of cascades is relatively hard ($\Gamma \sim 2$; see \citealt{ven10} and Section \ref{sec:results} of this paper), we do not expect substantial modifications of FSRQ spectra due to EM cascades (conversely, we do not expect a blazar with a hard measured spectrum to have a much softer intrinsic spectrum). Notably, the spectra of local universe FSRQs appear to be similar to that of the FSRQ population at large~\citep{agnlat10}. On the other hand, the spectra of BL Lacs are sufficiently hard that one might be concerned that the spectra are substantially modified by EM cascades. Nevertheless, we note that the population of local universe BL Lacs appear, on average, to have \textit{harder} spectra than the BL Lac population at large. As such, rather than attempting a likelihood analysis on the sparse local universe BL Lac population (which also have large measurement uncertainties in their spectral indices), we make use of the SID for the whole BL Lac population as this represents a more conservative choice. 

Finally, while the use of broken power-law spectra is motivated by our fits to the gamma-ray spectra of other types of blazars in the LAT energy range~\citep{vp11}, we should note that typical one-zone Synchrotron Self-Compton models of BL Lacs predict their gamma-ray spectra to be more curved (e.g., log-parabola). If the gamma-ray spectra of BL Lacs are more curved than the spectra selected here, one might expect there to be less cascade than predicted here since there would be fewer high-energy photons. On the other hand, the high-energy index beyond the break is already pretty soft ($\sim 3$) in our template spectra, so we do not expect the amount of cascade to change substantially if we selected log-parabola rather than broken power-laws. However, we acknowledge that the overall choice of template spectra (e.g., low-energy and high-energy indices and break energy) could affect the amount of cascade emission, and its impact on our results is currently being studied in more detail.

Since the AGN population peaks $\sim 1$, we take the typical redshift of blazars to be $z_{\rm src} = 1$. In modeling the suppression of the source brightness due to magnetic deflection of cascades (see Appendix), we take the jet opening angle of blazars to be $\sim 1^{\circ}$. The model of the EBL was taken from \citet{fra08}.

We also include a model of the SF galaxy contribution to the EGB. In this case, we assume that the gamma-ray luminosity of an SF galaxy is mainly due to the decay of neutral pions created through galactic cosmic-ray interactions with neutral hydrogen in the galaxy. It is thought that galactic cosmic rays are accelerated in supernova remnants and then diffuse throughout the galaxy. Hence, the flux of galactic cosmic rays is expected to be proportional to the galaxy's supernova rate, which is, in turn, proportional to its star-formation rate. Furthermore, the neutral hydrogen gas provides the fuel for forming stars. As such, the expectation is that an SF galaxy's gamma-ray luminosity can be parameterized in terms of its star-formation rate \citep{pf02,starblat10,fie10,mak11,sv11,sfgallat12}. For the purposes of this paper, we model the SF galaxy contribution using the infrared luminosity function model of \citet{sv11}. In this model, the star-formation rate of an SF galaxy is assumed to be proportional to its infrared luminosity; thus, the gamma-ray luminosity of an SF galaxy is proportional to a power of its infrared luminosity. In \citet{sv11}, the relationship between the gamma-ray luminosity of an SF galaxy and its infrared luminosity was determined by fitting the measured gamma-ray luminosities of {\it Fermi}-detected SF galaxies to their infrared luminosities. The SF galaxy contribution could then be determined by integrating over an infrared luminosity function of SF galaxies, which they took from \citet{hop10}.

\subsection{Electromagnetic Cascades from Blazars}\label{subsec:cascade}

The cascade intensity is given by
\beq\label{eqn:cascadeintens}
I^c_E = \frac{d^4N^c}{dtdAd\Omega dE}\,,
\eeq
where $dN^c/dE$ is the spectrum of cascade photons due to pair production and IC scattering \citep[for full equations and derivations, see][]{ven10}:
\begin{widetext}
\beq\label{eqn:cascadespec}
\frac{dN^c}{dE_0}(E_0) = \! \int_0^{z_{\rm max}} \!\!\! \int_{E_{p,{\rm min}}}^{E_{p,{\rm max}}} \!\! \int_{0}^{1} (1+z) \frac{d^2N_\gamma}{dzdE_p} P(f;E_p,z) \left[\frac{dN_{\Gamma_1}(E_0(1+z))}{dE}+\frac{dN_{\Gamma_2}(E_0(1+z))}{dE}\right] e^{-\tau(E_0,z)} df dE_p dz\,,
\eeq
\end{widetext}
\noindent where $dN_\Gamma/dE$ is the spectrum of IC scattered radiation per electron of Lorentz factor, $\Gamma$, $P(f;E_p,z)$ is the probability that the pair production interaction at a given redshift, $z$, of a primary photon of energy $E_p$ will produce electron-type particles of energies $E_{e1} = f\times E_p$ and $E_{e2} = (1-f)\times E_p$, $\Gamma_1 = f\times E_p/mc^2$, $\Gamma_2 = (1-f)\times E_p/mc^2$, and $d^2N_\gamma/dzdE_p$ is the continuous spectrum of photons undergoing pair production interactions as a function of redshift and primary energy. \\
\indent The full numerical integration of Equation \ref{eqn:cascadespec} is computationally cumbersome. As such, in order to calculate the cascade spectrum, one typically makes use of a Monte Carlo code, such as {\it Cascata} \citep{ven10}, that propagates photons from known spectra and redshift distributions and calculates the cascade spectrum of each photon. However, for the purposes of this paper, we employ a more semi-analytical model of the cascade development. Since we have assumed a typical redshift for all blazars, we have eliminated the need for randomly generated source redshifts. Instead, the spectrum of pair-producing photons, $d^2N_\gamma/dzdE_p$, can readily be determined from the \emph{absorbed} collective intensity of unresolved blazars binned in energy:
\vskip-0.3in
\begin{widetext}
\beq\label{eqn:abs-decoupled}
\int^{E_{0,{\rm max}}}_{E_{0,{\rm min}}} I^{\rm abs}_E(E_0)dE_0 = 
I_0 \int^{E_{0,{\rm max}}}_{E_{0,{\rm min}}} \!\! \int_{-\infty}^{\infty} d\alpha \, p(\alpha)\frac{\left[(E_0/E_b)^{\alpha_1}+(E_0/E_b)^{\alpha_2}\right]^{-1}}{\mathcal{S}(E_f,\alpha)}\left(1 - e^{-\tau(E_0,z)}\right)dE_0\,.
\eeq
\end{widetext}
\noindent For each energy bin, we determined a characteristic redshift for the start of the cascades based on the criterion that the characteristic redshift is that at which the number of photons that have been ``absorbed'' in propagating from the source redshift is half the number of the total photons ``absorbed'' in propagating to the present epoch. Using this criterion, the characteristic redshift can be calculated numerically from
\beq
\tau_{\ast}(E_0,z_{\ast}) = \ln\left[\frac{1}{2}\left(e^{\tau(E_0,z_{\rm src})}+1\right)\right]\,.
\eeq
Also, we assume that each secondary electron produced in a pair-production interaction carries half of the energy of the primary photon. Finally, since the number density of CMB photons is much higher than those of the other wavelengths that comprise the EBL, we assume that the seed photons for the IC interactions are CMB photons. In so doing, we can neglect Klein-Nishina effects in determining the spectrum of scattered radiation. The spectrum of scattered radiation per electron of Lorentz factor, $\Gamma$ is given by \citep[]{ven10}
\beq
\frac{dN_{\Gamma}}{d\epsilon_1} = \int \!\! \frac{d^3N_{\Gamma'}}{dtd\epsilon_1d\epsilon}dtd\epsilon = \int \!\! \frac{d^3N_{\Gamma'}}{dtd\epsilon_1d\epsilon}\left|\frac{dt}{dE_e}\right|dE_ed\epsilon\,,
\eeq
where $\Gamma'$ is the electron Lorentz factor at time, $t$, $\epsilon$ is the initial soft photon energy, $\epsilon_1$ is the scattered photon energy, $d^3N_{\Gamma'}/{dtd\epsilon_1d\epsilon}$ is the differential scattered photon spectrum given by \citep{blu70}
\beq
\frac{d^3N_{\Gamma'}}{{dtd\epsilon_1d\epsilon}} = \frac{3\sigma_{\rm T}c}{16\Gamma'^4}\frac{n(\epsilon)}{\epsilon^2}\left(2\epsilon_1\ln\frac{\epsilon_1}{4\Gamma'^2\epsilon} + \epsilon_1 + 4\Gamma'^2\epsilon - \frac{\epsilon_1^2}{2\Gamma'^2\epsilon}\right)\,,
\eeq
$n(\epsilon)d\epsilon$ is the number density of soft photons, and $| dE_e/dt |$ is the rate of energy loss of the electron given by
\beq
\frac{dE_e}{dt} = -\frac{4}{3}\sigma_{\rm T}c\Gamma'^2\int^{\epsilon_{\rm max}}_{\epsilon_{\rm min}} \!\!\!\!\! \epsilon n(\epsilon) d\epsilon\,,
\eeq
We note that this formalism assumes continuous energy losses for the electrons, which is appropriate for the regime that we are considering. Energy losses due to synchrotron emission are also included, but are negligible compared with IC losses for the magnetic field strengths considered. Furthermore, this simplified approach has the advantage that recycling cascades (in which IC photons up-scattered to high energies are also allowed to initiate EM cascades) is more feasible than in the full Monte Carlo approach. As such, for BL Lacs, we also calculated the spectrum of second generation cascades. Notably, the effect of EM cascades is to redistribute high-energy radiation to lower energies; as such, the high-energy component of second generation cascades necessary to produce higher generation cascades is greatly reduced. Thus, the contribution from generations beyond the second generation is negligible. For FSRQs, the intrinsic spectra of the sources are sufficiently soft that even the first generation of cascades makes a small contribution to their overall collective intensity.\footnotemark\ The suppression of the source brightness due to magnetic deflection of cascades (see Appendix) is also calculated, and we assume a uniform magnetic field that evolves solely due to the expansion of the Universe [$B(z) \sim B_0(1+z)^2$; \citealt{ner09}], where $B_0 \sim 5 \times 10^{-14}$ G. We should also note that for the purposes of this paper, we assume that the correlation length of the magnetic field is much larger than electron cooling distances. Alternative scenarios for the magnetic field and source spectra will be explored in future publications.

\footnotetext[4]{We note that the small amount of cascade radiation is not simply the result of the low break energy chosen for this analysis; \citet{ven10} modeled FSRQ spectra as unbroken power laws up to $\sim 100$ TeV and also found that the FSRQ cascade contribution was small.}

\subsubsection{A Note About Plasma Instabilities}\label{subsubsec:plasmabeam}

Regarding the stability of EM cascades during propagation, \citet{bro12} present the very interesting notion that plasma interactions between cascades and the intergalactic medium would introduce energy losses that are more significant than IC energy losses, thereby quenching the cascades. Unfortunately, this idea was premised on the assertion that cascades constitute plasma beams, even though the connection with plasmas is not obvious despite cascades containing charged particles. In order for a collection of particles to constitute a plasma, the particles must be able to influence each other electromagnetically (for this reason, the particles in a plasma must necessarily be charged particles). From our understanding of relativistic beams of particles, the electromagnetic interactions between particles are largely suppressed at very high energies~\citep{cha99}. We can see this by considering the field around a single charged particle at rest and then boosting in the direction of propagation. For large Lorentz factors (such as those involved in the EM cascade calculations), length contraction compresses the electric and associated magnetic fields so that they exist only in a plane perpendicular to the direction of propagation of the particle. Since the bulk of the particles in the cascades will not lie in this plane, they will not be electromagnetically connected to the particle in question. For the few particles that could lie in this plane, the Lorentz forces exerted on them largely cancel because they are relativistic and propagate co-linearly resulting in near cancelation between the electric and magnetic field components. For simple charge distributions (e.g., the two-dimensional Gaussian distribution), analytic expressions have been derived for the Lorentz forces on particles in a beam due to other particles within the beam, and such forces have been demonstrated to be suppressed by a factor of $\gamma^{-2}$~\citep{cha99}. As such, even for electrons with energies $\sim$ few MeV, the strengths of the forces exerted on them are suppressed to the percent level; at energies $\sim$ hundreds of MeV and above, these forces are negligible\footnotemark.

\footnotetext[5]{For this reason, it has been noted that collective effects are also negligible for high-energy charged particle beams such as those made in the Large Hadron Collider at CERN.} 

Another method for assessing the impact of collective behavior on pair cascades begins with calculating the \emph{relativistic Debye length for a pair cascade}, 
\begin{equation}\label{eqn:relDebye}
l_{\rm Debye}=\left(\frac{\gamma^2 k_{\rm B}T_{\rm cas}}{4\pi n e^2}\right)^{1/2},
\end{equation}
where $T_{\rm cas}$ is the transverse temperature of the cascade as measured in the laboratory frame (the frame of the intergalactic medium) and $n$ is the particle number density of the cascade. The expression for the cascade transverse temperature is given by \citep{bro12}
\begin{equation}\label{eqn:beamtemp}
\frac{k_{\rm B}T_{\rm cas}}{m_e c^2} \simeq \frac{p_{\parallel}}{2 m_e c^2}\left(\frac{p_{\perp}}{p_{\parallel}}\right)^2 \simeq 5 \times 10^{-7}\left(\frac{E_{\gamma}}{\rm TeV}\right)^{-1},
\end{equation}
where $p_{\parallel}$ and $p_{\perp}$ are the longitudinal and transverse momenta, respectively, and $E_{\gamma}$ is the energy of the primary photon. For the particle number density, we use the expression given by \citet{min12},
\begin{equation}\label{eqn:beamdens}
n \simeq 3 \times 10^{-25} \ {\rm cm}^{-3}\left(\frac{E_{\gamma}L_{\gamma}}{10^{45}\ {\rm ergs}/{\rm s}}\right) \left(\frac{D}{\rm Gpc}\right)^{-2}\left(1+z\right)^{-4},
\end{equation}
for which we take the mean free path for pair production for a TeV photon to be $0.8$ Gpc at $z = 1$ rather than using the full expression. Taking the reference values and plugging the results of Eqns. \ref{eqn:beamtemp} and \ref{eqn:beamdens} into Eqn. \ref{eqn:relDebye}, we find $l_{\rm Debye}$ to be $\sim 1$ kpc. 
For comparison, we can calculate the radius of the cascade, $a \sim \delta D_e$, where $\delta \sim m_e/E_e \sim \gamma$ is the angular dispersion of the cascade in the absence of magnetic fields, $D_e$ is the electron cooling distance assuming that IC scattering of CMB photons is the dominant cooling process \citep{ner09}\footnotemark,
\begin{equation}
D_e = \frac{3m_e^2c^3}{4\sigma_{\rm T}U_{\rm CMB}E_e} \simeq 10^{23}(1+z)^{-4}\left(\frac{E_e}{10\mbox{ TeV}}\right)^{-1} \mbox{ cm}\,,
\end{equation}
and we assume the electron energy is about half of the primary photon energy, $E_e = E_{\gamma}/2$. Assuming a primary photon energy of $1$ TeV and $z = 1$, we find the radius of the cascade to be $\sim 0.04$ pc. Thus, the Debye length of the cascade is much larger than its radius, so charge screening is ineffective, and the dynamics of the cascade are dominated by single-particle behavior rather than collective effects \citep{rei94}. Since the mean interaction rate for IC scattering of CMB photons is much larger than that for, say, bremsstrahlung interactions with protons in the intergalactic medium \citep{pro95}, we conclude that the process that dominates the cooling of pairs in cascades is indeed IC scattering. 

In the case of the nonzero IGMF, the compression of the electric and magnetic fields to a plane still applies. Due to magnetic deflection, the density of charges falls, suppressing the likelihood of finding nearby charges and further increasing the Debye length.
Therefore, we conclude that cascades do not constitute the necessary collective phenomena for substantial plasma interactions. 

\footnotetext[6]{Note that since the bulk of the cascade is produced in the first generation, we consider only one generation for this simple calculation.}

\subsection{Anisotropy Energy Spectrum}\label{subsec:aes}

The anisotropy energy spectrum, $C_\ell(E)$, is defined as the intensity fluctuation angular power (in units of sr) at a given angular scale, $\ell$, as a function of energy.  In the simple case of a two-component background, the total angular power is given by \citep{sp09}
\beq\label{full}
C_\ell^{\rm tot} = f_1^2C_\ell^{\left(1\right)} +  f_2^2C_\ell^{\left(2\right)} + 2f_1f_2C_\ell^{\left(1 \times 2\right)}\,,
\eeq
where $C_\ell^{\left(n\right)}$ is the angular power spectrum of component $\left(n\right)$, $f_n = I_n(E)/I_{\rm tot}(E)$ is the energy-dependent fraction of the total emission arising from component $(n)$, and $C_\ell^{\left(n \times m\right)}$ is the cross-correlation term for components $\left(n\right)$ and $\left(m\right)$. As $C_\ell^{\left(n\right)}$ is a measure of angular fluctuations in units of the mean, it is independent of energy for a single population of sources with identical spectra. If the two components are uncorrelated, $C_\ell^{\left(n \times m\right)} = 0$, and Equation~(\ref{full}) is further simplified, 
\beq\label{simple}
C_\ell^{\rm tot} = f_1^2C_\ell^{\left(1\right)} +  f_2^2C_\ell^{\left(2\right)}\,.
\eeq
If additional components {\em of negligible angular power} contribute to the total background signal, then these will not result in additional terms in Equation~(\ref{simple}), but rather they will affect $C_\ell^{\rm tot}(E)$ by changing the fractional contributions, $f$, of each of the components that {\em do} contribute to the angular power. 

In the simple scenarios we explore, we always consider two components that do contribute to the total angular power: BL Lacs and FSRQs. Because in our adopted model the two contributions do not add up to the total EGB signal measured by {\it Fermi} \citep{lat10a}, we assume that the rest of the EGB signal originates from a very low-$C_\ell$ population, such as SF galaxies. For simplicity, we take $C_\ell^{\left(\rm SF\right)} = 0$ for this component; hence, their contribution only enters Equation (\ref{simple}) through the total background intensity, $I_{\rm tot}(E)$ (i.e., the denominators of $f_1$ and $f_2$). As discussed in the previous sections, the energy spectrum of the primary emission from FSRQs is very soft at very high energies, so the cascade signal from FSRQs is negligible; all cascade emission that we consider comes from BL Lac VHE photons. 

In the limit of low $B$, there is little angular spread in the cascade photons, so these photons appear to originate from the same sources as the primary emission. In this case, the only effect of the cascades is to alter the energy spectra of blazars, leaving their anisotropy properties unchanged. 
As such, the fractional components of Equation~(\ref{simple}) become $f_{\rm bl}(E) = \left[I^{\rm int}_{\rm bl}(E) + I^{\rm casc}_{\rm bl}(E)\right]/I_{\rm tot}(E)$, while $C_\ell^{\left({\rm bl}\right)}$ remains unchanged.

\begin{figure*}
\begin{center}
\subfigure[]{\label{fig:1a}\includegraphics[scale=0.5, clip]{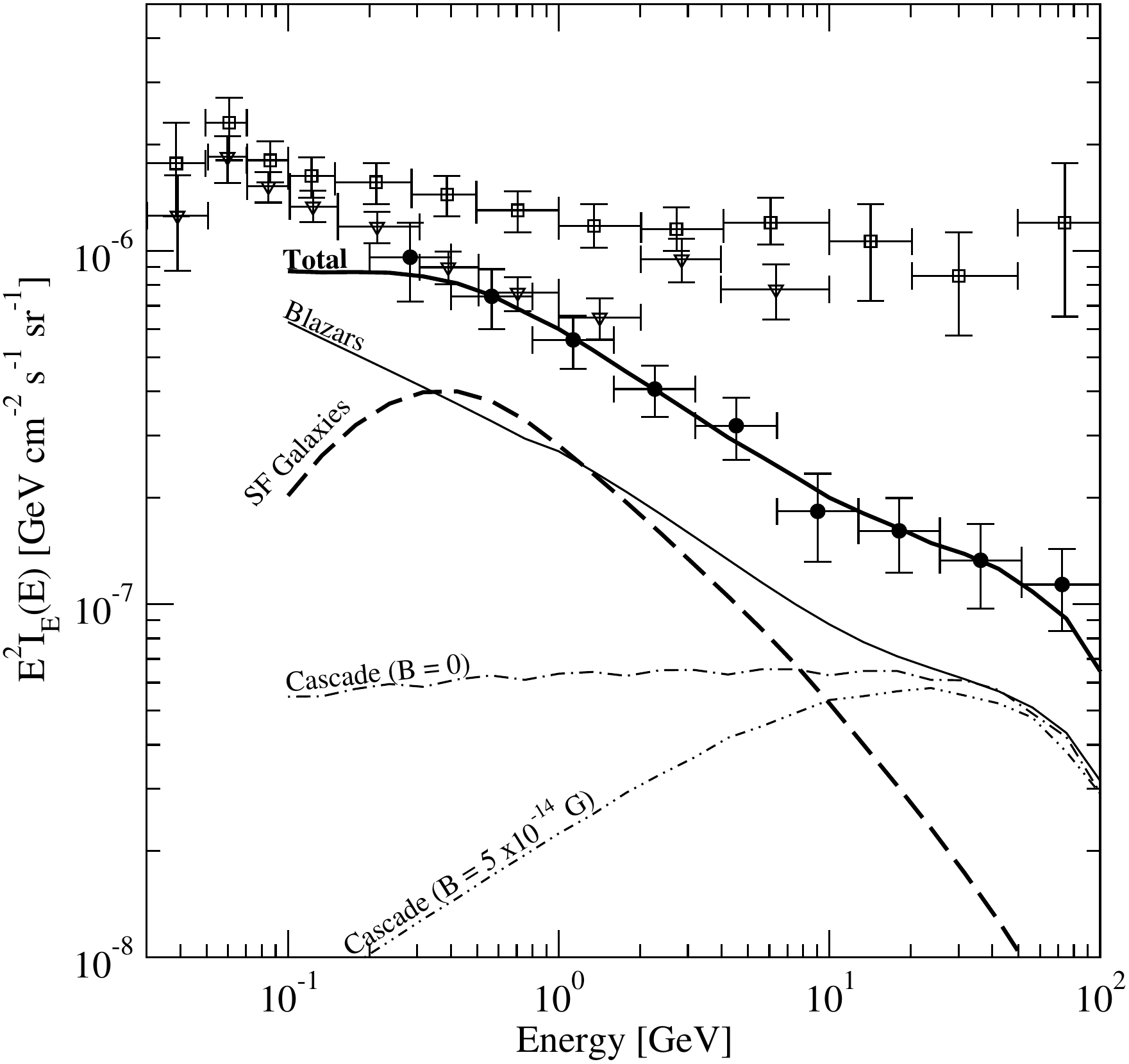}}\subfigure[]{\label{fig:1b}\includegraphics[scale=0.5,clip]{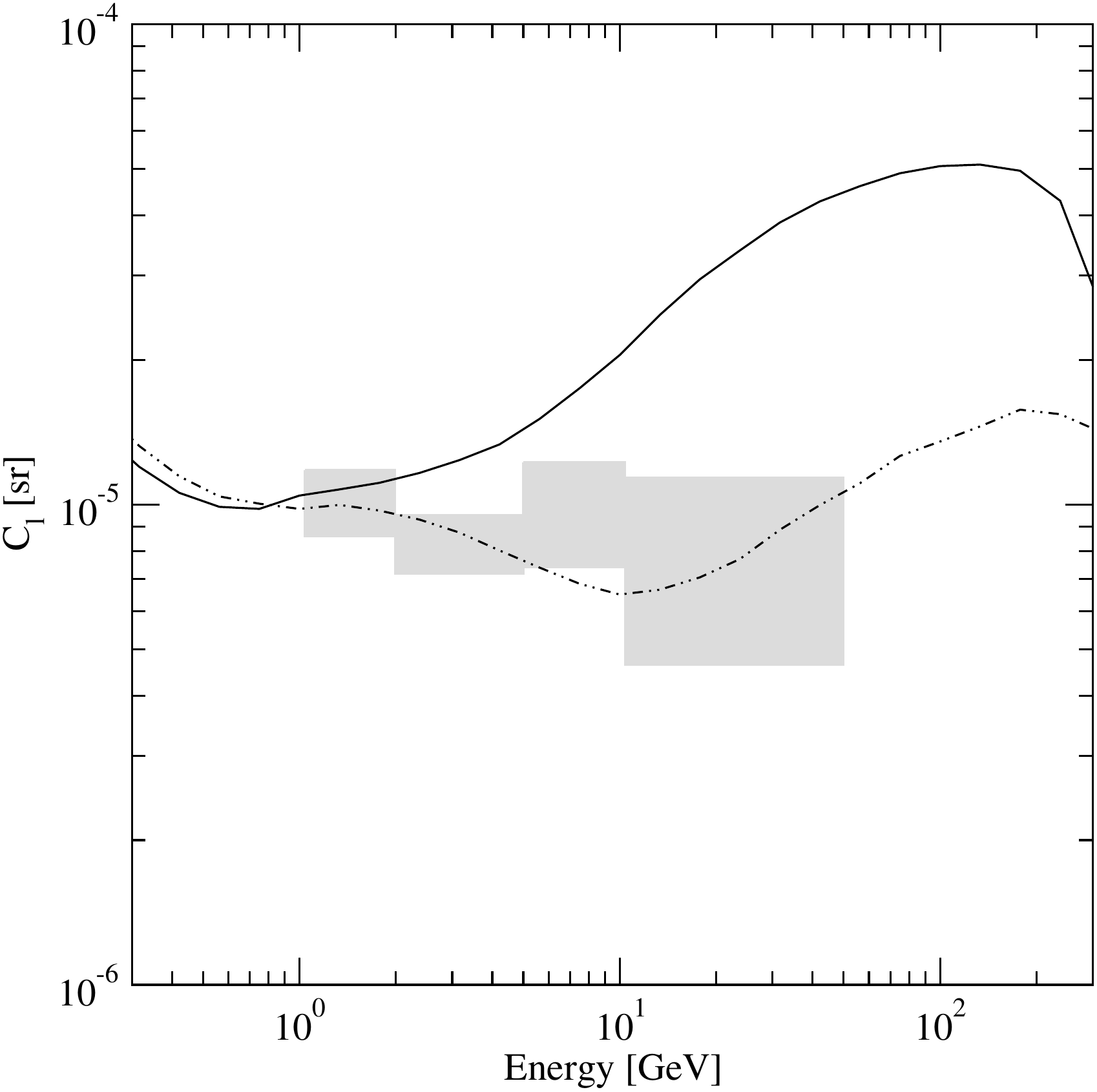}}
\caption{{\it a}: The spectrum of the modeled EGB (thick solid line) together with individual components: intrinsic blazar emission, including both FSRQs and BL Lacs (thin solid line); SF galaxies (dashed line); electromagnetic cascades for zero IGMF (dot-dashed line) and nonzero IGMF (double dot-dashed). For reference, the spectra of the EGB based on both {\it Fermi} (filled circles; \protect\citealt{lat10a}) and EGRET data (open squares; \protect\citealt{sre98}; open triangles; \protect\citealt{str04}) are also plotted. {\it b}: The anisotropy energy spectra of the modeled contributions to the EGB. The solid line is the total model assuming zero IGMF. The double dot-dashed line is the total model assuming nonzero IGMF. Grey boxes represent the {\it Fermi}-LAT measurement of the anisotropy energy spectrum of the EGB~\citep{anisolat12}.}
\end{center}
\end{figure*}

In the limit of strong $B$ ($B_0 \sim \mbox{few} \times 10^{-14}$ G), we determined the scattering angles of the cascades (see Appendix) and found that for observed photon energies less than $\sim  100$ GeV, the scattering angles were sufficiently large such that within the angular spread of the cascade emission on the sky (halo radius $\sim$ several degrees at $E_{\gamma} \sim$ tens of GeV), there would be enough sources to result in overlapping (confused) halos\footnotemark. Under these conditions, the cascade emission from the entire population of BL Lacs would be fairly isotropic, so we take the cascade signal as having angular power, $C_\ell^{\left({\rm casc}\right)} = 0$. Since the cascades now have an angular power that is different from the blazars and the null IGMF case, they will enter Equation (\ref{simple}) in the same manner as the SF galaxies, only through the total background intensity, $I_{\rm tot}(E)$. The numerators of $f_1$ and $f_2$ in Equation~(\ref{simple}) will then only include the intrinsic emission from FSRQs and BL Lacs, respectively (i.e., $f_{\rm bl}(E) = I^{\rm int}_{\rm bl}(E)/I_{\rm tot}(E)$). We emphasize that it is for \textit{this} reason that the total angular anisotropy energy spectra of the two magnetic field scenarios will be different. Even if the overall EGB cascade contributions in the two scenarios were the same, the total angular anisotropy energy spectra would be distinct, depending on the prominence of the cascades in relation to the other contributions to the EGB.



\footnotetext[7]{As based on the number of sources above the {\it Fermi}-LAT sensitivity above $10$ GeV. Since the blazar contribution to the EGB is the collective emission of \emph{unresolved} blazars, this criterion is quite conservative.}

Since, as stated in Section \ref{subsec:blazars}, a reliable BL Lac GLF has yet to be constructed, a full calculation of the angular power of BL Lacs as per \citet{and07} is not yet feasible. Instead, we normalize our anisotropy energy spectra such that in both scenarios, $C_\ell^{\rm tot} \sim 10^{-5} {\rm \, sr}$ around $1$ GeV, consistent with the {\it Fermi} measurement of angular anisotropies in the EGB at these energies \citep{anisolat12}. This results in a blazar angular power\footnotemark\ of $C_\ell^{\left( {\rm FSRQ}\right)}  = C_\ell^{\left({\rm BL \, Lac}\right)} = 6\times 10^{-5} {\rm \, sr}$, which will be testable in future publications once a reliable BL Lac GLF is found. The measurement at $1$ GeV was chosen to balance the competing needs of adequate angular resolution and sufficient photon statistics, and to minimize possible contamination from any remaining galactic foreground emission. The uncertainty at $1$ GeV is relatively small, though more data from {\it Fermi} will likely improve it.
\footnotetext[8]{Note that for a Poisson distribution of sources, the angular power at a given energy is constant with multipole.}

\section{Results}\label{sec:results}

In Figure \ref{fig:1a}, we plot the spectra of the modeled contributions to the EGB. The ``total'' contribution is the sum of the contributions from blazars, SF galaxies, and one of the cascade models (the total is similar for both models since the largest differences in the cascade contributions in the two scenarios occur at energies at which the EGB is dominated by blazars and SF galaxies). The blazar component hardens above $\sim$ few GeV due to the increased contribution of BL Lacs with respect to FSRQs, as BL Lac spectra remain hard while FSRQs soften. The cascade spectrum in the case of the zero IGMF is, as expected, quite flat ($dN/dE \propto E^{-2}$; \citealp[see e.g.,][]{str74,ven10}) demonstrating the cascade effect of redistributing radiation from high energies to lower energies. The sum of the three components fits quite well the spectrum of the EGB as measured by {\it Fermi}. Notably, the spectra of the cascades for the two cases of the IGMF, while similar at the highest energies, are quite dissimilar at the lower energies. This is due to the loss of some cascade radiation in the strong IGMF case as significant amounts of the cascades are deflected away from the line of sight of the observer. However, it should be noted that given our simple model of the blazar contribution to the EGB, we were only able to account for cascade losses due to deflection. That is, in our simple picture, we are calculating the amount of radiation that an observer looking head-on at a given source would be able to detect. It is likely that just as cascades from one aligned source are deflected out of the observer's line of sight, cascades from other aligned sources and even misaligned sources are deflected \emph{into} the observer's line of sight. Nevertheless, we note that even if the strong IGMF cascade radiation were comparable to that of the zero IGMF case, the angular power in the two cases would remain distinct since the angular properties of the cascades are different (see Section~\ref{subsec:aes}). \\
\indent In Figure \ref{fig:1b}, we plot the anisotropy energy spectra of the EGB for the two cases of the IGMF. Notably, there is an appreciable difference between the two cases starting at $\sim$ few GeV. This is due to the fact that above these energies, the contribution from EM cascades relative to the other components is much more significant than at lower energies. As a result, the impact of the isotropization of the cascades by the IGMF is much more significant at these energies. Aiding in the distinction is that the cascades \emph{augment} the anisotropy in the case of zero IGMF, because they enhance the contribution of the most anisotropic component. For comparison, the {\it Fermi}-LAT measurement of the anisotropy energy spectrum of the EGB~\citep{anisolat12} is also plotted. With just two years of data, the {\it Fermi} measurements seem to already favor the higher IGMF case over the null IGMF case. 
%

\section{Conclusions \& Discussion}\label{sec:concl}

We have studied the effect of a strong IGMF on EM cascades from blazars and the resulting impact on the cascade component of the EGB and the anisotropy energy spectrum of the EGB. We have shown that while the spectrum of cascade radiation in the case of zero IGMF is flat in $E^2$, the spectrum in the case of a strong IGMF is suppressed at low energies, though there could be an additional contribution arising from cascades from other sources, such as radio galaxies and propagating UHECRs. 


For the two extreme cases of IGMF strength we have considered here, we have calculated the anisotropy energy spectrum of the EGB. In the case of zero IGMF, the effect of cascades is to \emph{augment} the anisotropy arising from blazars since the cascades increase the contribution to the EGB arising from blazars and are highly collimated with the primary emission. In the case of the strong IGMF, the cascades are deflected substantially and the angular properties of the parent blazar population are blurred. Thus, in the case of a strong IGMF, the effect of cascades is to \emph{reduce} the anisotropy of the EGB as they hinder the anisotropy arising from blazars rather than reinforcing it. At energies at which the relative contribution to the EGB arising from EM cascades is significant, the anisotropy energy spectra of the two cases are distinguishable, and current {\it Fermi} data already seem to favor non-negligible IGMF values. With more data and further refinement of this technique, {\it Fermi} will be able to provide tighter constraints and possibly even {\it measure} the IGMF.

It is worth noting that if blazars and their corresponding cascades make a large contribution to the {\it Fermi} EGB spectrum above $1$ GeV, then the null IGMF scenario would be in tension with current {\it Fermi} anisotropy measurements (see Figure \ref{fig:1b}). Similar observations have been made by \citet{cuo12} and \citet{har12}, leading both groups to conclude that blazars can explain no more than $20\%$ of the EGB above $1$ GeV lest they overproduce the EGB anisotropy. However, our results demonstrate that in the scenario of a non-zero IGMF, it is possible that blazars and their corresponding cascades can contribute substantially to the {\it Fermi} EGB spectrum above $1$ GeV and even reproduce the EGB spectrum above $10$ GeV \emph{without overproducing the EGB anisotropy}.

Many possible contributions to the EGB have been identified in the literature, and a multi-component background is plausible in light of current {\it Fermi} data. The extent to which other contributions will impact the difference in anisotropy energy spectra in the two cases considered here depends on their angular power and whether they also have appreciable cascade contributions. For instance, other gamma-ray--emitting radio galaxies for which the jet is not aligned with the observer's line of sight (``misaligned'' blazars) could also contribute to the EGB \citep{ino11}. For those radio galaxies that are too misaligned for the observer to receive substantial intrinsic gamma-ray emission, a nonzero IGMF could deflect cascades into the observer's field-of-view, giving rise to an additional contribution to the EGB, provided that the spectra of these sources are hard enough to initiate cascades. These sources will be substantially fainter than blazars, but are not likely to be as numerous as SF galaxies. Hence, they are not likely to have substantial angular power and will enter the calculation of the total angular power of the EGB in a manner similar to that of the SF galaxies, solely by virtue of their relative contribution to the \textit{intensity} of the EGB. We will return to the impact of radio galaxies in a future publication.

In addition to that from radio galaxies, there could also be a contribution to the EGB from the annihilation of dark matter in the Galaxy. Even if gamma rays from dark matter annihilation do not dominate the EGB in any energy band, they could have a substantial impact on the total angular power of the EGB provided that they make a significant contribution to the EGB intensity. The impact of dark matter annihilation on both the intensity and angular power of the EGB depends on the mass of the dark matter particle and the combined cross section for annihilation into channels that result in gamma rays. Some models of the EGB contribution from dark matter can result in spectra that peak at the same energies at which the EM cascades contribute most significantly to the EGB. However, the significance of the dark matter contribution to the EGB also depends on the amount of substructure present in the dark matter halo with the largest signals occurring in scenarios in which the halo contains numerous large clumps~\citep{sie08}. In such scenarios, the angular power of dark matter is substantial and will likely enhance the total angular power of the EGB at the energies at which the dark matter signal peaks \citep{sp09}. In both of the IGMF cases considered in this paper, the impact of a substantial contribution from clumpy dark matter at energies of tens of GeV would be to increase the angular power in the last energy bin in Figure \ref{fig:1b}. The angular power in the two scenarios would still be distinct, though the distinction would depend on the significance of the dark matter contribution relative to that of blazars and EM cascades. Nonetheless, in order to push the two models to yield the same angular power in the last energy bin, the contribution from dark matter would have to be substantial, and even in the case of nonzero IGMF, it could prove difficult to explain the flatness in total angular power of the EGB with energy as measured by the \textit{Fermi}-LAT~\citep{anisolat12}. Even worse is the scenario in which there is no substantial cascade contribution, but there is a substantial dark matter contribution -- without a low-anisotropy contribution to mitigate the total angular power, the angular power in the last energy bin would rise substantially. We must also consider the possibility that the dark matter halo is smooth rather than clumpy. In this case, the angular power of dark matter would be minimal, but it could be difficult for dark matter to contribute significantly to the EGB. In any case, \textit{Fermi} has already begun to constrain models weakly interacting massive particles as dark matter candidates, including tantalizing hints of a signal at $\sim 130$ GeV near the Galactic Center \citep[see e.g.,][]{wen12}. In the future, the much anticipated results from the Large Hadron Collider could provide even more clues on the nature of dark matter. Thus, the question of the dark matter contribution to the EGB could soon be answered.

Finally, there could also be an appreciable cascade contribution from the interactions of UHECRs with the EBL. The impact of cascades from UHECRs on the total angular power of the EGB depends on the still-unknown sources of UHECRs, but the most likely scenario is that the difference in angular power between the two IGMF cases will still be present and may, in fact, be augmented. In order for the separation in angular power to disappear with the inclusion of UHECR cascades, the UHECR sources would have to have very little angular power. Such a scenario might have been possible in the top-down models of UHECRs, but such models are tightly constrained by the upper limits on the UHE photon flux determined by the Pierre Auger Observatory, and some such models have already been excluded~\citep{aug09}. As such, the current discussion of the origins of UHECRs largely plays out in the astrophysical arena, and the classes of astrophysical sources that are typically thought to be capable of accelerating UHECRs tend to be rare and powerful sources (as are blazars). Hence, the scenario in which UHECR sources exhibit very little angular power is unlikely. Furthermore, for electrons with energies greater than those considered here, the IC process becomes more inefficient as Klein-Nishina effects become important. Therefore, the cascade photons from UHECR interactions that would contribute to the EGB are likely produced by electrons with energies similar to those found in the VHE gamma-ray cascades produced here. Furthermore, UHECRs are themselves \textit{charged} particles and would have already been deflected by the IGMF prior to initiating cascades. Thus, the inclusion of UHECR cascades would likely also result in a difference in angular power between the two IGMF cases, which may even be more prominent than demonstrated in this paper. Given the possibility of a large contribution to the EGB from UHECR cascades~\citep[e.g.,][]{ahl10} and the possible implications for the sources of UHECRs, we will devote a future publication to the study of the impact of UHECR cascades on the angular power of the EGB.

We note that an advantage of our approach is that rather than focusing on a few select sources, our method employs information from the whole observable sky. As such, our approach could prove capable of providing more sensitive constraints on the \emph{global} properties of the IGMF. Furthermore, our approach is not limited to uniform magnetic fields with large coherent lengths; other field configurations can be probed with this technique. In practice, in order to use this technique to provide an IGMF measurement, intermediate cases of IGMF strength need to be considered as well, and the uncertainties entering the calculation of the expected anisotropy energy spectra in each case carefully evaluated. In the case of intermediate IGMF strengths, a numerical simulation of the expected angular power is appropriate as the angular anisotropy properties of the cascade radiation will be altered with respect to the primary BL Lac emission, though not completely suppressed as in the case of the strong IGMF. In fact, since the cascade radiation is \emph{correlated} with the primary blazar emission, the cross-correlation term of Equation~(\ref{full}) would have to be calculated. As such, for intermediate IGMF strengths, a simple analytical calculation of the cascade effect on the total angular power is not possible. Concerning uncertainties, these enter our calculation primarily through the model assumptions for each EGB component (which can, in turn, self-consistently determine their anisotropy properties), and through the EBL model. We will expand our investigation of the practical usage of the anisotropy energy spectrum of the EGB to constrain the properties of the IGMF in future publications.

In the meantime, this simple, proof-of-concept calculation has already demonstrated that the anisotropy energy spectrum can be a powerful tool in constraining the IGMF. With a few more years of data from {\it Fermi}, we can begin to probe the IGMF and uncover more clues into the origins of UHECRs and large-scale structure formation.

\section*{Acknowledgments}

We thank Jennifer Siegal-Gaskins and Brandon Hensley for helpful discussions about anisotropies in the EGB,  Kostas Tassis for helpful discussions about astrophysical magnetic fields, and Amy Lien, Brian Fields, and Floyd Stecker for helpful discussions about astrophysical contributions to the EGB. We also thank the {\it Fermi}-LAT team, especially the members at NASA Goddard Space Flight Center, for many helpful discussions about the {\it Fermi}-LAT detector characteristics and {\it Fermi} data. We thank the anonymous referee for constructive comments that helped to improve the manuscript. TMV would like to thank the Caltech Astronomy Department for its hospitality during her visits to Pasadena while working on this and other projects.

\bibliography{ms_igmf_short_bibtex}
\bibliographystyle{mn2e}

\appendix

\begin{figure*}\label{fig:2}
\begin{center}
\subfigure[]{\label{fig:app1a}\includegraphics[scale=0.5]{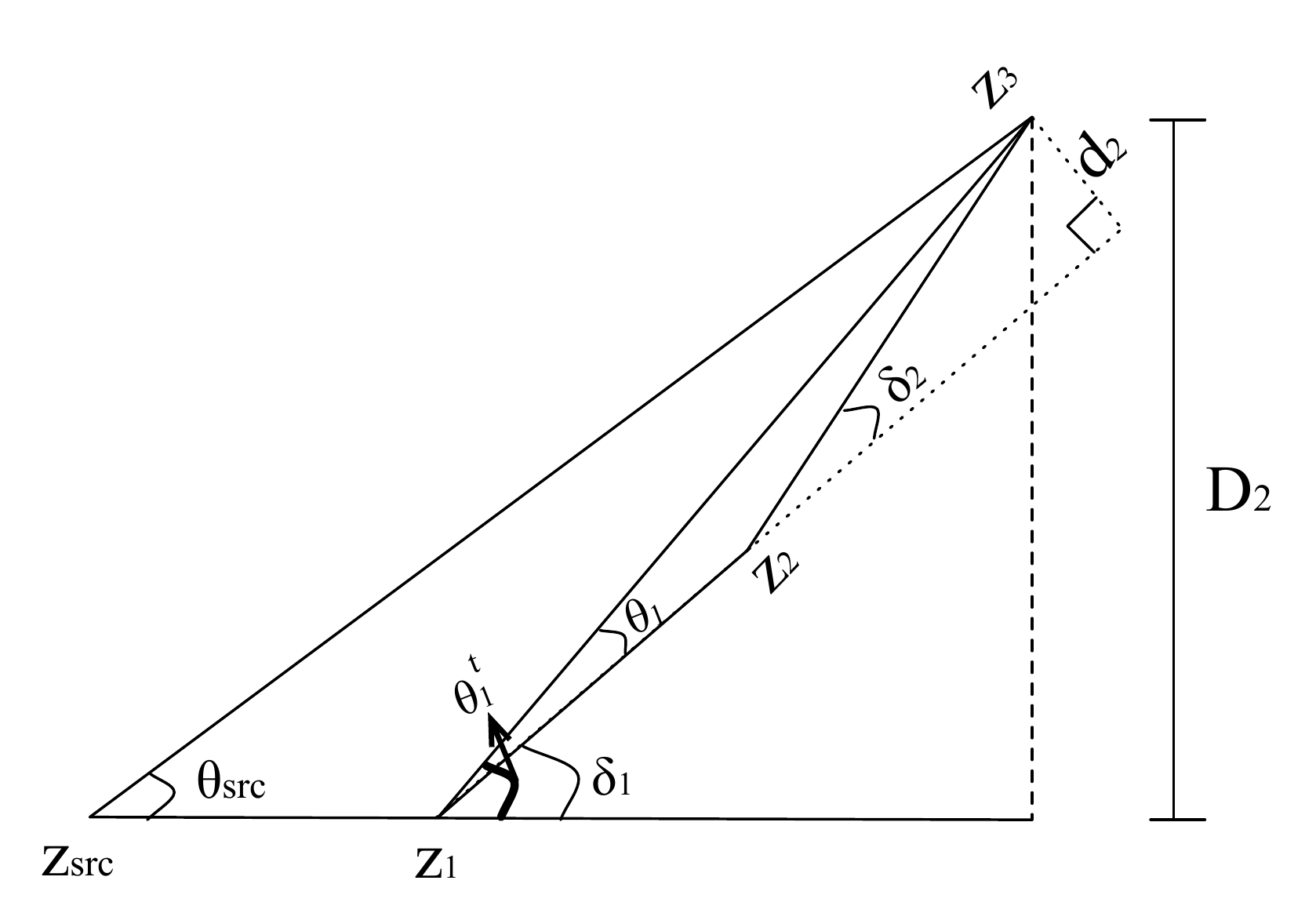}}\subfigure[]{\label{fig:app1b}\includegraphics[scale=0.5]{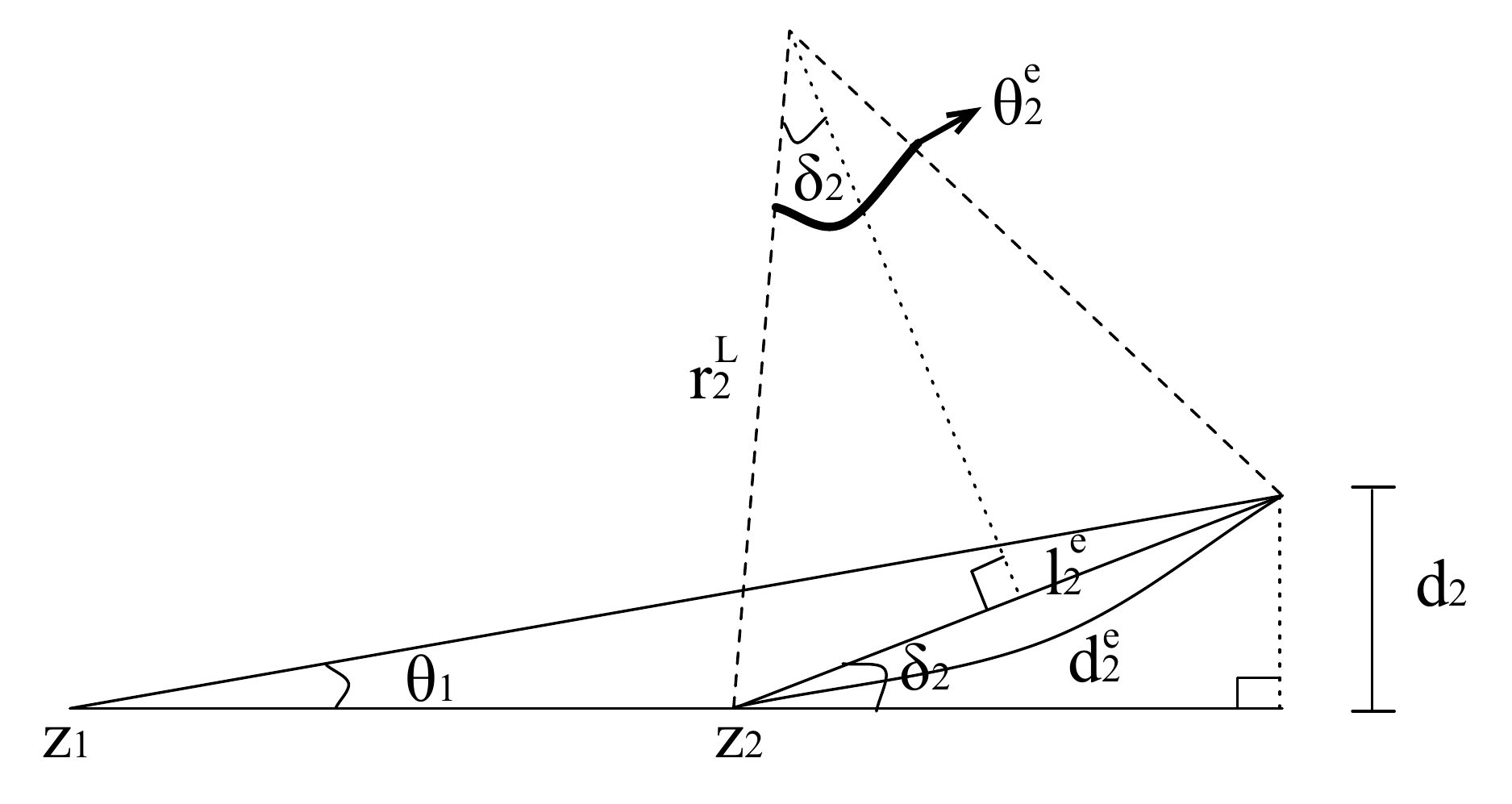}}
\caption{{\it a}: Geometry for the angular deflection of two generations of EM cascades. {\it b}: Geometry for the angular deflection of one generation of EM cascades.}
\end{center}
\end{figure*}

\section{Suppression of the Source Brightness Due to Magnetic Deflection of EM Cascades}\label{app:suppression}

In calculating the suppression of the source brightness due to magnetic deflection of EM cascades, we must determine the angular spread of the cascades, $\theta_{\rm src}$, as viewed by a fundamental observer at the source redshift, $z_{\rm src}$ (Figure A\ref{fig:app1a}). In order to determine $\theta_{\rm src}$ for multiple cascades, we begin with considering the problem for one generation of cascades (Figure A\ref{fig:app1b}). In this case, $\theta_1$, the angular spread of the cascade as viewed by a fundamental observer at $z_1$, can be expressed in terms of quantities evaluated at the redshift of interaction of the ``primary'' photon, $z_2$, and the angular diameter distance from $z_{\rm src} = z_1$ to $z_2$, $D_A(z_1,z_2)$:
\beq\label{eqn:d2divDAz1z2}
\theta_1 = \frac{d_2}{D_A(z_1,z_2)}\,,
\eeq
where $d_2$ is the proper lateral distance the electron has traveled as measured by a fundamental observer at $z_2$ (see e.g., \citealt{lon08}), and we neglect Klein-Nishina effects, so the electron cooling distance, $d^e_2$, is not cosmological. By geometry, $d_2$ is given by
\beq
d_2 = l^e_2\sin\delta_2 = 2r^L_2\sin^2\delta_2\,,
\eeq
where $l^e_2$ is the straight path of $d^e_2$, $r^L_2$ is the Larmor radius at $z_2$ of an electron with energy, $E^e_2$, in a magnetic field of strength, $B(z_2) = B_0(1+z_2)^2$, given by \citep{ner09}
\beq
r^L_2 = (3 \times 10^{28} \, {\rm cm})\left[\frac{B(z_2)}{10^{-18}\ {\rm G}}\right]^{-1}\left(\frac{E^e_2}{10\ {\rm TeV}}\right)\,,
\eeq
and $\delta_2 = (1/2)\, \theta^e_2$ is the angle through which the electron is deflected away from the straight-line path of the ``primary'' photon, which is equal to half of the magnetic deflection angle of the electron given by \citep{ner09}
\beq\label{eqn:thetae}
\theta^e_2 = \frac{d^e_2}{r^L_2} = (3 \times 10^{-6})\, (1+z_2)^{-2}\left(\frac{B_0}{10^{-18} \ {\rm G}}\right)\left(\frac{E^e_2}{10\ {\rm TeV}}\right)^{-2}\,.
\eeq
The angular diameter distance from $z_1$ to $z_2$ can be found from the angular diameter distance from $z_2$ to $z_1$ through the Reciprocity Theorem:
\beq
D_A(z_1,z_2) = \frac{1+z_1}{1+z_2}D_A(z_2,z_1)\,,
\eeq
where $D_A(z_2,z_1)$ for a flat universe is given by
\beq
D_A(z_2,z_1) = \frac{1}{1+z_1}\left[D_c(z_1)-D_c(z_2)\right]\,,
\eeq
and $D_c(z)$ is the comoving distance at $z$.

Given $\theta_1$ in terms of known quantities, we can find $\theta_{\rm src}$ for two generations of cascades using an equation similar to Equation \ref{eqn:d2divDAz1z2}. Returning to Figure A\ref{fig:app1a}, we note that in this case, the primary photon is emitted at $z_{\rm src}$ and interacts at $z_1$ emitting electron-type particles that do not propagate cosmological distances. One electron-type particle is deflected away from the straight-line path of the primary particle through an angle, $\delta_1$, and up-scatters CMB photons. The up-scattered photons then propagate until they interact at $z_2$, creating more electron-type particles. Since we neglect generations of cascades higher than the second generation, $\theta_1$ is given by the previously developed formalism. The angle through which the electron at $z_1$ is deflected away from the straight-line path of the primary photon, $\delta_1$, is also given by Equation \ref{eqn:thetae}, but for quantities evaluated at $z_1$.

Given $\theta_1$ and $\delta_1$, the total angular spread of the cascades as viewed by a fundamental observer at $z_1$ is $\theta^t_1 = \theta_1+\delta_1$. The lateral spread of the cascade as measured by a fundamental observer at $z_2$, $D_2$, is given by Equation \ref{eqn:d2divDAz1z2}:
\beq
D_2 = \theta^t_1D_A(z_1,z_2)\,.
\eeq
Finally, we can again use Equation \ref{eqn:d2divDAz1z2} to find the angular spread of the cascades as viewed by a fundamental observer at the source redshift:
\beq
\theta_{\rm src} = \frac{D_2}{D_A(z_{\rm src},z_2)}\,.
\eeq
The flux suppression of the cascade due to magnetic deflection can be found from the flux of the cascade in the absence of magnetic fields by relating the solid angle of the deflected radiation (determined from $\theta_{\rm src}$), $\Delta\Omega_d$, to that of the unaffected emission (in this case, the solid angle of the jet), $\Delta\Omega$:
\beq
F_{\Delta\Omega_d} = \frac{\Delta\Omega}{\Delta\Omega_d}F_{\Delta\Omega}\,.
\eeq

We note that this procedure is only valid for non-isotropically emitting sources. Conceptually, there should be no flux suppression for an isotropic source because the symmetry of the problem demands that whatever emission that is deflected out of the observer's cone of sight is replaced by emission being deflected in.

\bsp

\label{lastpage}

\end{document}